# ON THE POSSIBILITY OF NEW HIGH $T_c$ SUPERCONDUCTORS BY PRODUCING METAL HETEROSTRUCTURES AS IN THE CUPRATE PEROVSKITES

Antonio Bianconi

*University of Roma, Department of Physics, P. A. Moro 2, 00185 Roma, Italy*

Here we propose a possible way to produce superconductors with high critical temperature $T_c$ by confinement of a Fermi liquid in a superlattice of quantum stripes (wells, wires or dots). The enhancement of $T_c$ is obtained by tuning the size L of the stripes (wells, wires or dots), and the wavelength of electrons at the Fermi level $\lambda_F$ at the resonance condition $L \sim \lambda_F$. The 3D superconducting phase is stabilized by the condition that the separation W between the stripes (wells, wires or dots) should be of the order of the coherence length $\xi_0$. This quantum state is realized in $Bi_2Sr_2Ca_2Cu_2O_{8+y}$ superconductor with $T_c = 84K$, where the resonance condition $k_{Fy} \sim 2\pi/L$ for the component of the Fermi wavevector normal to the stripe direction holds. This work gives the parameters for the design of new superconducting materials with higher $T_c$ made of metal heterostructures approaching the atomic limit.

The search for superconductors with a high critical temperature $T_c$ [1] has been a long standing subject in solid state physics. The extended efforts to produce high $T_c$ metallic materials failed to raise the critical temperature above 23 K [2] which seems to be the saturation value for metallic superconductors. The breakthrough was realized by Müller and Bednorz [3] by turning the direction of the research from metals to doped ceramic materials. New high $T_c$ superconductors with the critical temperature in the range of 20-150K have been found by synthesizing doped cuprate oxide perovskites, with a record of $T_c \sim 150K$ in $HgBa_2Ca_2Cu_3O_{8+y}$ [4]. The mechanism driving the superconducting state from the range $0 < T_c < 23K$ of metals and alloys to the high temperature range $20 < T_c < 150K$ of doped cuprate oxide perovskites, i.e. enhancing critical temperature by a factor ~10, is not known.

**The main barrier to understand the quantum state of the electron gas in cuprate perovskites is the difficulty to determine their structure**. In fact, these materials show a two dimensional conductivity in the $CuO_2$ planes. The average long range structure of the $CuO_2$ planes is well known but the short range structure diverges from the average crystallographic

structure [5,6]. In fact the Cu site local structure configurations are non homogeneous and their actual distribution is not established. In a long term project to investigate the local structure of the $CuO_2$ planes by Extended X-ray Absorption Fine Structure (EXAFS) we have found that the $CuO_2$ plane in the $Bi_2Sr_2CaCu_2O_{8+y}$ (Bi2212) is not homogeneous. The Cu sites show a landscape of structural configurations that has been called with the nick name "**the corrugated iron-foil landscape**" to stress the introduction of a one-dimensional character in the 2D electron gas [7].

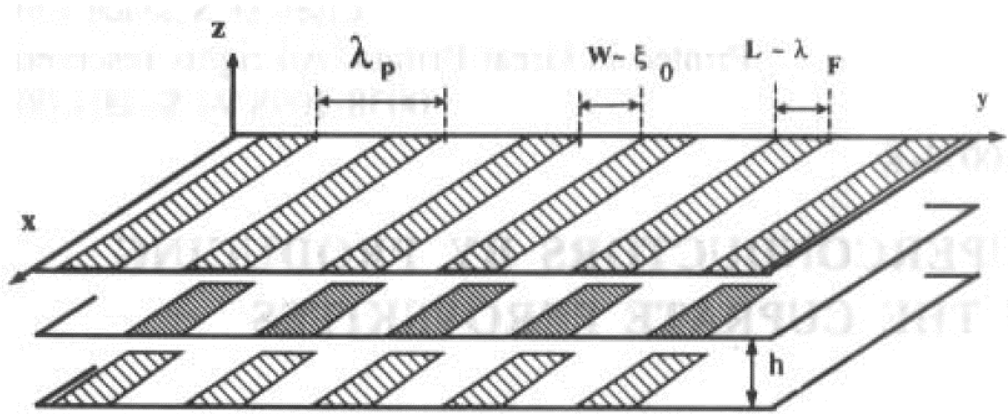

**Fig. 1** The superlattice of quantum stripes in the xy planes where a superconducting electron gas is confined under the conditions that $L \sim \lambda_F$, $W = \xi_0$ and the spacing h between the planes should be also of the order of $\xi_0$ to realize the 3D superconducting phase. Here we show that this quantum state is realized in the cuprate high $T_c$ superconductors.

The $CuO_2$ plane is decorated by two types of parallel stripes: 1) stripes of width L locus of the Fermi liquid, where the Cu site structure configurations is of the type of the low temperature orthorhombic (LTO) phase of $La_2CuO_4$ with a long Cu-O(apical) bond; and 2) a second type of stripes, called here the barriers, of width W locus of the polaronic charges, where the distorted Cu site configurations are characterized by a short Cu-O(apical) bond and a tilting of the pyramids in the ($\pi 0$) direction as in the Low Temperature Tetragonal (LTT) phase of $La_{1.875}Ba_{0.125}CuO_4$ [8]. The holes injected in the barriers, associated with polarons, are expected to be localized at T=0K because they will be in the same quantum state as in insulating $La_{1.875}Ba_{0.125}CuO_4$ at the critical density number of holes per Cu site $\delta_c = 1/8$ [9,10]. The 2D-

*key words: new materials by design, heterostructure at the atomic limit, shape resonance*

Fermi liquid is therefore confined in a superlattice of quantum stripes of width L as shown in Fig. 1. The parallel stripes in the **x** direction form a superlattice in the **y** direction with period $\lambda_p$.

The local hole (electron) density in the stripes forming the Fermi liquid with **the "large" Fermi surface** shown in Fig. 2a is $1+\delta_i$ ($n_i=1-\delta_i$) where $\delta_i$ is the local hole number per Cu site. The local hole density in the barriers is $\delta_l = \delta_c=1/8$. The Fermi surface in the barriers is assumed here to be formed by **a "small" Fermi surface** as in $La_{1.875}Ba_{0.125}CuO_4$ [10] as shown in Fig. 2b.

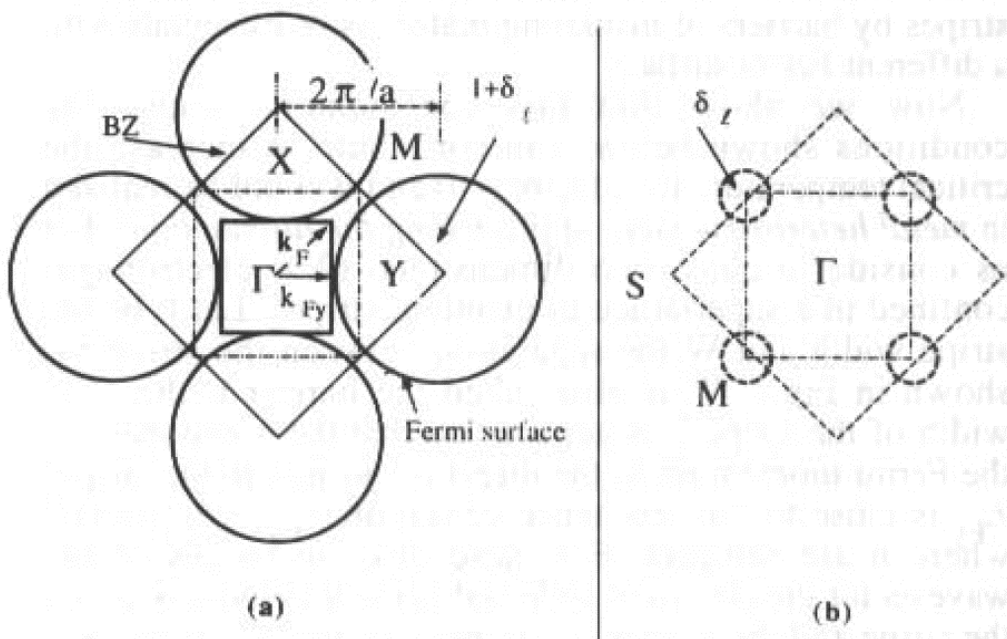

**Fig. 2** Pictorial view of, panel a) the large Fermi surface for the Fermi liquid localized in the stripes of undistorted lattice (LTO) and long Cu-O(apical) distance and panel b) the small Fermi surface formed by a small hole pocket for localized holes in the barriers with the LTT like structure and short Cu-O(apical) bond with critical doping 1/8. The small pocket is centered at the M point in Bi2212 because of the mixing with the BiO band but it is expected at the S point where it is determined by the $CuO_2$ plane only as predicted by Pickett et al. [10].

The electronic structure of this system is like that of a metal heterostructure where a metallic material is confined in the stripes by barriers of insulating materials or by metals with a different Fermi surface. Now we show that this confinement, under the conditions shown below, can contribute to increase the critical temperature by a factor ~10 and it could be realized in *metal heterostructures approaching the atomic limit*.

Let us consider a **quasi two dimensional (2D) electron gas confined in a superlattice of**

*key words: new materials by design, heterostructure at the atomic limit, shape resonance*

**quantum stripes**. Let L be the stripe width and W the separation between the stripes as shown in Fig. 1. W is also called the barrier width. The width of the stripe L is adjusted so that the component of the Fermi momentum in the direction normal to the stripes $k_{Fy}$ is close to the resonance condition $k_{Fy} \sim k_{ny} = n\pi/L$ where n are integers. For these discrete values of the wavevector the electrons, reflected at the interfaces between the stripe and the barrier, are trapped inside the stripe for a localization time depending on the probability for hopping between the stripes of the superlattice [11].

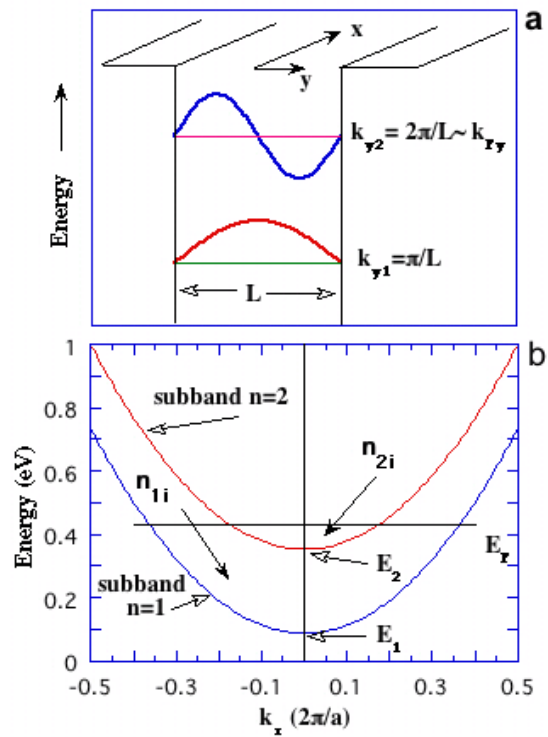

**Fig. 3 a)** The wavefunctions of the electrons at the n=1 and the n=2 resonances in a single stripe with an infinite barrier. **b)** The energy dispersion for two subbands with quantized $k_{y1}=\pi/L$ and $k_{y2}=2\pi/L$ for an ideal single quantum stripe as function of the momentum component $k_X$ in the direction of the free electron motion along the stripes, with an effective mass m*=2. $k_X$ is given in units of $2\pi/a$, where a=5.4 Å. The case where the Fermi energy is close to the bottom of the second subband is shown.

These resonances are present both for electrons in the discrete and in the continuum part of the energy spectrum and are similar to the multiple scattering resonances for electrons, called shape resonances, usually observed in the x-ray absorption near edge structure (XANES) [12] where electrons of energies as high as 20 eV are localized in the real space at the resonance condition. These resonances have been recently observed in metallic quantum wells [13].

Let us consider the particular simple case of a single stripe of width L as shown in Fig.3a. The electronic structure will be formed by subbands of quantized wavevector $k_{ny} = n\pi/L$ as shown in Fig. 3b. The number of occupied subbands is given by n= Int $[2L/\lambda_F]$. If the Fermi energy $E_F$ is close to the bottom of the second subband as shown in Fig.3b, only two main subbands will cross the Fermi level and the density of states can be separated into two partial density of states associated with the first $n_{1i}$ and the second $n_{2i}$ subband. The Fermi energy will be at the 2nd sharp maximum of the nearly one-dimensional density of states. The electrons in the second subband will be close to the resonance condition with $k_F \sim k_{Fy} \sim 2\pi/L$. In the superlattice the electrons will have a dispersion in the **y** direction determined by the superlattice subband and the hopping between the stripes.

In the Bi2212 monocrystal where the 2D-electron gas is confined in the superlattice of quantum stripes as in Fig. 1 in order to establish how many subbands are occupied and if the Fermi energy is at the resonance condition we have to know the stripe width and the Fermi wavevector.

We have measured the period of the superstructure $\lambda_p \sim 4.7a = 25.4$ Å (where a=5.4 Å is the lattice constant of the $CuO_2$ square lattice with Cu-Cu distance d=a/$\sqrt{2}$) in the direction of the **b** axis (**y** direction) the width of the barrier W= 2a= 10.8 Å and the width of the stripes L= 2.7a= 14.6 Å in a single phase Bi2212 monocrystal, with $T_c$= 84K [7].

Accurate measurements of the Fermi surface of Bi 2212 single crystal with $T_c$=85K have been reported [14]. The electron distribution curves show very sharp peaks in some directions of the Brillouin zone close to the Fermi level as in $YBa_2Cu_3O_7$ [15] that cannot be explained by the standard band structure calculations. The value of components of the Fermi wavevector $k_{Fx}$=$k_{Fy}$=0.38 (measured in units of $2\pi/a$) along the ΓM direction have been found, where a large superconducting gap opens up at $T_c$ [7].

Therefore we conclude that there are only two main occupied subbands Int $[2L/\lambda_F]$= Int [2 L $k_F/2\pi$] = 2 and that the component of the Fermi wavevector in the **y** direction is close to the

- 5 -

*key words: new materials by design, heterostructure at the atomic limit, shape resonance*

resonance condition

$$L\, k_{Fy}/2\pi = 2.7 \cdot 0.38 \sim 1.025 \pm 0.1 \qquad (4)$$

The photoemission data show anisotropy in the X and Y direction in agreement with the presence of stripes separated by a large energy barrier of width W=10.8 Å and small hopping between the stripes.

The density of states of the superlattice of quantum stripes [11] is dramatically different from the density of states of the 2D square lattice. Peaks several tens times larger than the corresponding 2D density of states appear and the superconducting critical temperature can be pushed up by a factor of the order 10 where the Fermi energy is tuned at one of these maxima. In fact for a standard superconducting metal following the BCS theory $T_c \sim 1.13\, \omega_D \exp(-1/N_0 V)$, where $N_0$ is the density of states at the Fermi energy and V is the electron-phonon coupling constant, therefore the increase of $N_0$ implies an increase of $T_c$. Band structure calculations of the cuprates give the electron-phonon coupling constant V~1.5 and the density of states $N_0$~0.15 states/eV-atom -spin showing that $N_0 V \sim 0.225$ in $La_2CuO_4$. Therefore by taking $\omega_D$~500K~43 meV as the Debey temperature for the Cu-O bonds measured by EXAFS [17] we can calculate in first approximation, the critical temperature of a homogeneous $CuO_2$ plane $T_c \sim 7$ K predicted by the BCS theory. More refined calculations of $T_c$ using the Allen-Dynes equation give $T_c$~30K [16].

**The enhancement of the critical temperature by forming metallic stripes of width L separated by stripes of width W can be calculated by following the solution of the gap equation of Thompson and Blatt [18] in a single film of a superconducting metal where the Fermi level is close and above the energy of the n resonance.** The enhancement factor at the second resonance as found in the cuprates should be of the order of $\exp(1/(3N_0 V))$~5 for a superlattice in comparison with the homogeneous $CuO_2$ plane. Therefore the critical temperature can be enhanced by the confinement from the 7K (20K) range to the 35K (100K) range.

We now propose that it should be possible to raise the critical temperature by producing metal heterostructures of standard superconducting materials by realizing the similar physical conditions that the electron-lattice instabilities have realized in the cuprate superconductors.

**Metal heterostructures approaching the atomic limit can be built by molecular beam epitaxy or similar modern technologies.** The 3D superconducting state is stabilized by a

superlattice [19] where the distance between the planes and the quantum stripes (wells, wires, or dots) is of the order of the superconducting coherence length $\xi_0$. In fact **in a superlattice is possible to raise $T_c$ by quantum confinement but in a single quantum well, as proposed by Thompson and Blatt, proximity effects and fluctuations [20] will suppress the superconducting phase and $T_c$.**

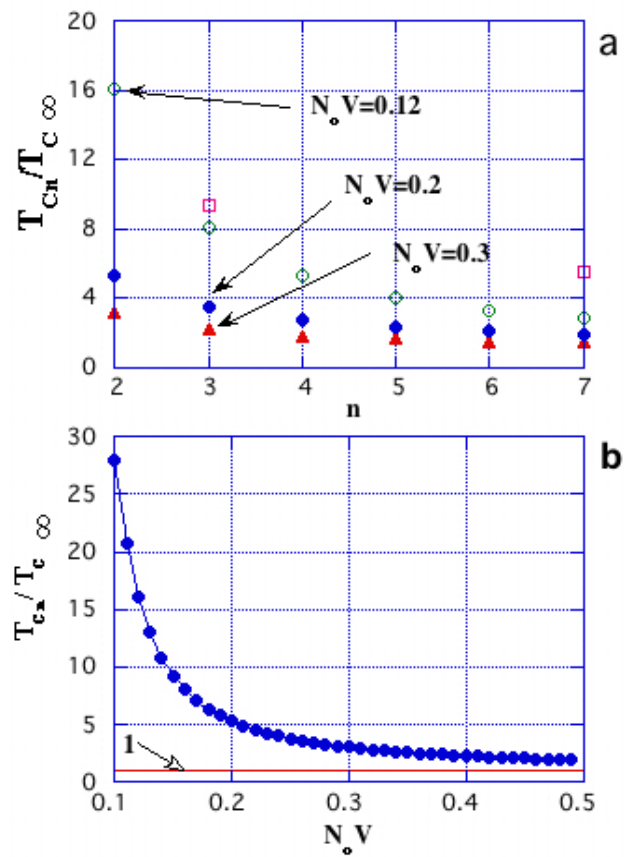

**Fig. 4** a) The ratio of the critical temperature $T_{cn}$ for a quantum well, with the Fermi energy tuned at the n resonance normalized to the bulk critical temperature $T_c$ for a superconducting metals with different coupling constants $N_oV$, calculated by using the Thompson and Blatt approach. The inverse of the threshold input energy $1/E_t$, in arbitrary units, as function of the resonance number n, $d=n\lambda/2$, in a microlaser measured by De Martini et al.[21] is shown by empty squares. **b)** The ratio $T_{cn}/T_c\infty$ as function of the coupling constant $N_oV$ for a stripe were the Fermi energy is tuned to the n=2 resonance.

*key words: new materials by design, heterostructure at the atomic limit, shape resonance*

The amplification of $T_c$ will be realized in metal heterostructures by tuning the width of the stripe (or the thickness of the wires, wells or dots) to fit the n=1, 2, 3 resonances $n = 2L/\lambda_F$. We have shown here that in the cuprate superconductors the n=2 resonance is realized. The amplification factor depends on the resonance number n and on the coupling term $N_0V$.

In Fig. 4a we report the enhancement factor for the case of a superlattice of quantum wells as function of the resonance number for the case of $N_0V$=0.3, 0.2 and 0.12, calculated by using the Thompson and Blatt approach. It is therefore clear that the largest amplification is obtained by tuning $E_F$ at the lowest resonances. The amplification for the superconductors as Nb, in the strong coupling regime $N_0V$=0.3, is expected to be of the order of 3 at the n=2 resonance and of the order of 6 at the n=1 resonance. In Fig. 4b we report the variation of the enhancement factor for the n=2 resonance as function of the coupling constant $N_0V$. It is clear that the largest amplification is obtained by going to superconductors in the weak coupling regime. Therefore it is clear that the high critical temperatures can be obtained by starting from superconducting materials with a large prefactor $\omega_D$ and in the weak coupling. This result sheds light on the fact the confinement increases the coupling term but the ultimate highest temperature that will be possible to reach depends on the energy of the virtual excitations exchanged in the pairing.

It seems to us that it is worth to investigate if in fullerenes the high critical temperature is due to a confinement of a Fermi liquid in a 3D-superlattice of quantum spheres. Finally it is interesting to remark that the quantum state of the electron gas in high $T_c$ superconductors is similar to the quantum state of the electromagnetic field confined in a microlaser under the conditions that the microcavity spacing $d=n\lambda/2$, [21] where n is an integer. In fact the inverse of the threshold of input energy versus microcavity order diverges going to the lowest n=1 resonance. We have reported the inverse of the threshold of input energy versus microcavity order measured by De Martini et al. in Fig. 5a and it clearly shows a similar dependence on the resonance number, as $T_{cn}/T_{C\infty}$.

In conclusion we have shown that the confinement of a superconducting metal in a superlattice of quantum stripes (wells, wires, spheres or dots) stabilizes the superconducting phase raising the critical temperature. In order to drive the cuprate superconductors to the highest $T_c$ allowed for a Fermi liquid with a coherence length $\xi_0$ of the order of $\lambda_F$ the microscopic pairing mechanism should be different from the electron-phonon process as in metallic low $T_C$ materials. In any case the quantum confinement should play a key role to stabilize the superconducting state at high temperature.

I am indebted to M. Missori and G. Raimondi for help in this work, and to C. Di Castro for many stimulating discussions.